\documentclass[12pt]{revtex4}
\usepackage{amsmath,amssymb}
\usepackage[mathscr]{euscript}
\usepackage{graphicx}

\DeclareMathAlphabet{\mathpzc}{OT1}{pzc}{m}{it}

\newcommand{\evalat}[3]{\left.#1\right|_{#2}^{#3}}
\newcounter{aaa}

{\end{trivlist}}
\newenvironment{teor*}[2][{}]{\begin{trivlist}%
\labelsep=0pt\item[\bfseries #2. ]#1}
{\end{trivlist}}

\DeclareMathOperator{\Bd}{Bd}
\newcounter{statm}
\newcommand*{\CauD}{\mathcal{D}}
\newcommand*{\field}{\mathsf{f}}
\newcommand*{\metr}{\mathsf{g}}
\newcommand*{\isom}{\phi}
\newcommand*{\por}{\lessdot}

\newcommand{\statmaph}[2]{\par\refstepcounter{statm}\textbf{\arabic{statm}. #1.}{\parindent=0pt #2\par\smallskip\par}}

\newcommand{\ssy}[5]{#1 \emph{#2} {\bf #3} (#4) #5\rlap{.}}

\begin{document}
\title{What is faster --- light or gravity?}
\author{S. Krasnikov\thanks{Email: S.V.Krasnikov@mail.ru}}
\affiliation{The Central Astronomical Observatory at Pulkovo}

\begin{abstract}
General relativity lacks the notion of the speed of gravity. This is inconvenient, and the current paper is aimed at filling this gap. To that end I introduce the concept of  the ``alternative" and argue that its variation called  the ``superluminal alternative" describes exactly what one understands by the ``superluminal gravitational signal". Another, closely related, object called the ``semi-superluminal alternative" corresponds to the situation in which a massive (and therefore gravitating) body reaches its destination sooner than a photon \emph{would}, if the latter were sent \emph{instead} of the body. I prove that in general relativity constrained by the condition that only globally hyperbolic spacetimes are allowed, 1) semi-superluminal alternatives are absent and 2) under some natural conditions and  conventions admissible superluminal alternatives are absent too.
\end{abstract}

\maketitle
%
 \section{Introduction}
 The goal of this paper is to compare the speed of gravity with the speed of light within general relativity. In this section we discuss a major obstacle in achieving this goal, which is the lack of a suitable --- that is physically motivated, but rigorous --- definition of the ``speed of gravity" in the general case (i. e., say, beyond the linearized theory). Without such a definition any answer to the question posed in the title of this paper is obviously meaningless, but the reason for its lack is quite valid: the Universe according to relativity is a ``motionless", ``unchanging" 4-dimensional object, and gravity is just its shape. But what can be called the speed of a  shape? What is the ``speed of being a ball"?

 Still, there are situations in which it would be convenient to be able to assign a speed to gravity, or at least to be able to tell whether it is greater/less than the speed of light.
Consider, for example, an observer   orbiting a red giant.
 Suppose one day the events $s$, $b$, and $a$ happen: $s$ is the star exploding as a supernova, $a$ is the observer seeing the explosion, and $b$ is the observer's equipment showing that the  local geometry has drastically changed as a result of the same explosion. The sought-for definitions must allow one to say that the propagation of gravity was    superluminal, if $b \prec a$ [written also $b\in I^{-}(a)$, or $a\in I^{+}(b)$], i.~e. if there is a piecewise timelike  future directed curve from $a$ to $b$. To put it slightly more mathematically let us write $x\por y$ for ``$x$ gravitationally affects $y$" or, interchangeably, ``$x$ is a (gravitational) cause of $y$". Then
our task is to define the ``gravitational cause" so that the gravitational signalling would be
 recognized superluminal if and only if there   is a pair
  of  points $s,b$
 such that $s\por b$  even though $s\not\preccurlyeq b $, where $s\preccurlyeq b $, or $s\in J^{-}(b)$ means ``there is a piecewise nonspacelike  future directed curve from $s$ to $b$".
  By saying so we, of course, have not   solved the problem, but  have made it clearer, or so it seems. All one needs now is  to  put forward an intuitively acceptable criterion for whether an event acts gravitationally upon another   (in the example with the supernova this fact was hidden in the words ``a \emph{result} of the same explosion").

  Probably the simplest  step along these lines is to declare that  $s\por b$ when and only when the two events can be connected by a piecewise smooth curve --- called \emph{gravitational signal} --- defined by a   condition imposed on its velocity. For example, the tangent to the curve might be required to be null, or to obey some constraints involving velocities often mentioned in discussing   superluminal signaling by material fields:  phase velocity, group velocity, or the velocity of transport of energy.  For the gravitational field, however, this approach does not work. In the general case  it is even hard to define those quantities, but there is also a more serious reason for the failure. Let us turn for a moment to material fields.
\statmaph{Example}{\label{ex:period} (I) Consider a Minkowski space with  a
field $\field$ in it which  obeys only the  equation
\begin{equation}\label{eq:EM}
\Box\, \field=\sum_{k=1}^{K}\delta(x-x_k - v_kt),
\end{equation}
where the  constants $K$, $x_k$, $v_k$ --- are free parameters of the theory.
It is not that simple to  justify  any particular definition of the signal   in this case  (hereafter we shall touch on
 that).  What \emph{is} clear in advance, however, is that  any \emph{reasonable} definition of signals must be satisfied, in particular, by the future directed    lightlike broken lines (otherwise one will have to reinterpret the entire Special relativity, with its thought experiments involving essentially the same  field).
Similarly, spacelike separated points must prove to be causally disconnected. So, for instance, the point $b_N$ with the coordinates $t=1$, $x=1$, $y=N$, $z=0$ is affected by the    origin of the coordinates, when $N=0$, but not when $N=1$.

It is noteworthy that such a choice of the cause-effect relation makes the interpretation of the lines $x=x_k + v_kt$ with $v_k < 1$ and with $v_k>1$  strikingly different. While the former geodesics are just the world lines of ordinary pointlike charges (the zero acceleration may mean that their masses are very large), the latter ones do not correspond to any particles \emph{at all}. Each point of such a line is causally disconnected from all others. So instead of propagation of a particle we have a \emph{process} which takes place \emph{independently} at every point of the line (cf.~a light spot  running along a remote surface \cite{Ugarov}) and which consists in $\field$ infinitely growing prior to such a point and falling immediately after its occurrence.

(II) Consider now a theory in which the field $\field$  obeys the wave equation, but also  is subject to an \emph{additional} condition of the periodicity in the $y$ direction:
\begin{equation}\label{eqLperiod}
\field(t,x,y+1,z) = \field(t,x.y.z).
\end{equation}
Now whatever information about the origin of coordinates $o$ can be inferred from the values at $b_0$  of the field and its derivatives ${\field,}_{\mu...}(b_0)$, exactly the same information will be available to an observer who measures ${\field,}_{\mu...}(b_N)$, $N\neq 0$.
 So we have to  conclude that in \emph{this} theory $o\por b_N\ \forall N$.
}

There are no reasons whatsoever to believe  in   periodic fields. The example is cited only to demonstrate that (i) Equations of motion alone cannot determine the causal structure of a theory. Correspondingly, none of the aforementioned velocities can serve as the signal speed and (ii)
Two events ($o$ and $b_2$, for instance) can be causally related ($o\por b_2$), even though they are not connected by a signal understood as a curve $\sigma(\tau)$ such that
 \[
 \sigma(0) =o,\qquad \sigma(1) =b_2 ,\qquad \sigma(\tau_1) \por \sigma(\tau_2) \quad\text{at }
 \tau_1 < \tau_2.
 \]
  In this sense the relation $\por$ is not quite local.

From the foregoing it appears that one ought to abandon (at this stage, at least) the concept of the signal and to define the relation $\por$  directly from its physical meaning, in the spirit of the preceding examples.
In all appearance the notion of ``cause" will be satisfactorily captured by a relation $\por$, if the latter has the following properties:
\begin{enumerate}\renewcommand{\theenumi}{P\arabic{enumi}}
  \item $\por$ is a partial order relation. Indeed, it must be transitive (since the cause of a cause is obviously a cause), reflective (it is just a matter of convention and we  choose the analogy with the relation $\preccurlyeq$), and antisymmetric  (an event different from $a$
      cannot be both a cause and an effect of $a$).
  \item if $a\por b$, then there exists a set $S$ such that\label{prop:2}
  \begin{enumerate}
    \item  $S$ determines $\field(b)$ in the sense that  the   values taken in $S$ by the field and its derivatives $ {\field,}_{\mu...}(x)$, $x\in S$ fix uniquely the value $\field(b)$;
    \item  if $A$ is a neighbourhood of   $a$, then $S-A$  does \emph{not} determine  $\field(b)$.
  \end{enumerate}
 \end{enumerate}
 The requirement \ref{prop:2}  is justified by the fact that it is  an embodiment of the  idea that
\begin{enumerate}
         \item[\ref{prop:2}*.] any change in the effect is produced only by a change in some of its causes.
       \end{enumerate}

 The relation $ \por  $ is not defined \emph{uniquely} by those properties; for example, the relation   $\gtrdot$ defined by the equivalence
 \[a \gtrdot b\quad \Leftrightarrow\quad  b\por a
 \]
 presumably also possesses them. To fix the non-uniqueness one may need an additional convention, which is not surprising: different definitions of the causal order within a given theory account for different  views on what is freely specifiable in that theory.

   One might wish the above formulated definition to be more strict, but by and large it seems adequate in discussing causal properties of matter fields. It could be expected that in the gravitational case the   cause-effect relation can be introduced in the same manner, one only must take $\field $ to be the metric.
Presumably, it is this conviction that suggests the following  simple resolution  of the problems considered in this paper: ``The solution [to the Einstein equations] obtained depends, at a point $x$, only on the initial data
within the hypercone of light rays [\dots] with vertex $x$, that is, on the relativistic
past of that point. This result confirms the relativistic causality principle as
well as the fact that gravitation propagates with the speed of light'' \cite{Bruhat}. The flaw in this resolution is that ``is fixed as a solution of a differential equation by the data within a set $S$" and ``is caused only by points of $S$" is  not the same. In other words, the causality relation in the gravitational case may not obey \ref{prop:2}. This is, in particular, because the principle   \ref{prop:2}* does not apply to the metric. The point is that while for a material field $\field$ it is quite clear what ``a change in  $\field(p)$" is, there is no such thing as  a ``change in the metric at $p$". Indeed, in considering a spacetime $(M_1,\metr_1)$ one can give a precise meaning  to the words ``the geometry of a   set $V_1\subset M_1$ has changed": they mean that we consider \emph{another} spacetime $(M_2,\metr_2)$ and state that   there is a set $V_2\subset M_2$ and an isometry $\phi$ which maps $M_1-V_1$ to $M_2-V_2 $, but which cannot be extended to an isometry mapping the entire $M_1$ to $M_2$. However, that change  cannot be resolved into pointwise changes: there is no way, in the general case, to put in correspondence a particular $p_2\in V_2$ to each $p_1\in V_1$ (note that $V_2$ even need not be diffeomorphic to $V_1$) so as to compare $\metr_2(p_2)$ to $\metr_1(p_1)$ and thus to find out whether the metric in $p_1$ has changed.

It is clear from the foregoing that there is no easy way of introducing the relation $\por$. Therefore we take a completely different approach.

\section{Alternatives}
In this section we  formulate conditions which being imposed on a   pair of spacetimes   $M_1$ and $M_2$   allow one to speak of that pair as describing  two different extensions of a \emph{common prehistory} (in the example which opens the paper this prehistory would include the life of the red giant prior to  the explosion $s$). That will enable us to translate
the question of whether relativity (in a broad sense) admits superluminality of any kind  into the question of when the difference between such   $M_1$ and $M_2$ is attributable   to a certain event and its consequences \cite{speed}.

 \statmaph{Definition}{\label{def:alt}
A pair of pointed inextendible spacetimes
$(M_k,\metr_k,s_k)$, $k=1,2$  is called  an
\emph{alternative}, if there exists a pair of open connected past sets  $N_k \supset \Bigr(J^-(s_k)- s_k\Bigl)$ and an isometry $\isom$ which maps $N_1$ to $N_2$ and $J^-(s_1)- s_1$ to $J^-(s_2)- s_2$  (all matter fields in $N_1$ and $N_2$ are assumed to be tensors related by the same $\isom$).
}
 \statmaph{Notation}{ For a given alternative the pair $N_1,\isom$
 need not  be unique. Let
$\{N^\alpha_1,\isom^\alpha\}$ be the family of all such pairs. By $(N^*_1,\isom^*)$
we shall denote its
\emph{maximal} element, that is   one  which is not ``smaller"   than any other:
\[
\nexists \alpha_0 \colon\qquad N^*_1\subsetneq N_1^{\alpha_0},\quad \isom^*=
\evalat{\isom^{\alpha_0}}{N^*_1}{}.
\]
Correspondingly, $N^*_2\equiv \isom^*(N^*_1)$.%
 }

 The  existence of  $(N^*_1,\isom^*)$ follows from Zorn's lemma, since the open subsets of $M_{1}$ and $M_{2}$ are partially ordered by inclusion
 \[A\leq B \ \Leftrightarrow\ A\subset B,
  \]
 and with such an ordering every chain $\ldots\leq A_1\leq A_2\leq\ldots$   has an upper bound  $\cup_i A_i \subset M_{1,2}$.

 \statmaph{Comment}{\label{com:alt}
It is  the regions $N^*_k\subset M_k,\  k=1,2$ that describe the mentioned prehistory. The requirement that they be isometric is self-obvious. It is also obvious why both of them must be past sets (two spacetimes evidently do not describe the same region of the Universe, if their inhabitants differ in remembrances). As was explained in the Introduction, our main interest is actually \emph{non}-isometric regions of $M_k$ and we need $N^*_k$ only as a tool for outlining those regions. That is why we   require $N^*_k$ to be connected and maximal. Finally, the points $s_k \in M_k$ describe  the event (the star explosion in the mentioned example) responsible for  splitting the  evolution of the Universe into the two branches.}

\statmaph{Definition}{
The sets
 $\EuScript F_k\equiv \Bd  N_k^*,\  k=1,2$ will be termed  \emph{fronts}. A front $\EuScript
 F_k$ is \emph{superluminal}, if $\EuScript F_k \not\subset
 \overline{J^+(s_k)}$.}
 Being the boundary of a past set a front is a closed, imbedded,
achronal three-dimensional $C^{1-}$ submanifold \cite[Proposition 6.3.1]{HawEl}. At either $k$ the front $\EuScript F_k$ bounds the region $M_k - \overline{N_k^*}$, in which, loosely speaking, the  remembrances  (concerning the gravitational or material fields) of every observer  differ from what they would remember, if some other event happened in $s$. We interpret such a difference as evidence that the mentioned observer received a signal from $s$. Correspondingly, when such an observer  is located out of $\overline{J^+(s_k)}$ the signal is superluminal, hence our definition.

The concept of an alternative is quite rough. In the general case it does not make it possible to assign a specific speed to a ``gravitational signal", if by the latter a front is understood. Even the source of the signal cannot be determined uniquely: the same pair of spacetimes can satisfy the definition of alternative with different choices of points $s_k$. Nevertheless, it allows one to formulate a \emph{necessary} condition for calling the speed of gravity superluminal. Namely, in considering a particular theory (i.~e. a set of material fields and their relation to the geometry of the spacetime) let us single out a class of \emph{admissible} alternatives, by which the alternatives are understood consisting of  spacetimes $M_{1,2} $ such that they are equally possible  in that theory and differ only by the events $s_k$ and by the events which we agree to recognize as consequences of $s_k$ (not as consequences of some primordial difference in the spacetimes).
If none of  the admissible  alternatives has a superluminal front we shall acknowledge that the speed of gravity in this theory is bounded by the speed of light.

\section{Superluminal   gravitational signals in GR}

Let us adopt the convention that an alternative $(M_k,\metr_k,s_k)$ in which both spacetimes are globally hyperbolic is admissible, only if there are
  Cauchy surfaces $\EuScript S_k\subset M_k$ such that
\[
s_k\in \EuScript S_k,\qquad
\EuScript S_2-s_2=\isom^*(\EuScript S_1-s_1)
\]
and the values of material fields (and their derivatives, if necessary) in any $p\in \EuScript S_1$ are the same as in  $\isom^*(p)$.  Such a criterion does not look far-fetched, for, if there is no such a pair of  Cauchy  surfaces, why should one regard the difference in
$M_1$ and $M_2$ as ensuing from what happened in $s$ and its consequences, cf.~\cite{Low}? It rather must be acknowledged as  \emph{primordial}.

The global hyperbolicity of $M_{1,2}$ implies the equality
\[
M_k- J^+(s_k) = [J^-(s_k)  - s_k]\cup\,\CauD (\EuScript S_k-s_k),
\]
where   $\CauD (X)$ denotes the Cauchy domain of the set $X\subset M_k$  (i.~e. the set of  all points $p$ of $M_k$ such that every inextendible nonspacelike curve through $p$ meets $X$).
 By the existence and uniqueness theorem
equality of the initial data fixed at initial 3-surfaces implies isometry of the corresponding  Cauchy domains.  So, if an alternative is admissible, $\CauD (\EuScript S_k-s_k)$ are isometric and hence  $N_k^*$ [which by definition include $J^-(s_k)- s_k$]  include also $M_k-
J^+(s_k)$. Thus, neither of the fronts is superluminal.
In this sense general relativity does prohibit superluminal propagation of the gravitational field: under the formulated above assumptions \emph{the speed of a gravitational signal does not exceed the speed of light}.

\statmaph{Remark}{ The approach developed in this paper is suitable for other geometric theories as well. For example, to analyze the signalling in a theory dealing only with the causal relations between events, not with the entire metric, it suffices to replace the words \emph{an isometry $\isom$} in definition~\ref{def:alt} by \emph{a conformal isometry $\isom$}. Likewise, one might be interested in a theory   which considers only Ricci flat (i.~e.\ empty, if the Einstein equations are imposed)  spacetimes. The proposition proven in \cite{Bon-Sen} and reformulated in terms of alternatives says that in such a theory superluminal alternatives  turn out to be prohibited, if  an alternative is admissible only when in one of its spacetimes the  Weyl tensor vanishes to the future from a Cauchy surface through $s$.
}

It is important that the above-mentioned existence and uniqueness theorem is proven only under some ``physically justified" assumptions   regarding the properties of the right hand side of  the Einstein equation. A possible set of such assumptions is listed, for example, in \cite{HawEl} and one of them is that the stress-energy tensor is at most a polynomial in $g^{ab}$ (the corresponding assumption in \cite{Wald_GR} allows the tensor to include also the first derivatives of the metric). But those assumptions are \emph{known} to fail in some physically
interesting situations. In particular, vacuum polarization typically leads to the appearance
 in the right hand
side of the Einstein equations of terms   containing  second derivatives of the metric. Which suggests that strong  seniclassical effects like those expected  in the early Universe, or near  black hole horizons,  may lead to superluminal propagation of gravity.

\section{``Semi-superluminal" alternatives}\label{subsec: poluFTL}

The fact that  a single event is associated with two fronts, either in its own spacetime, has a quite non-trivial consequence because they do not need to be superluminal both \emph{at once}.

\statmaph{Definition}{ An alternative is called \emph{superluminal} if both its fronts are superluminal and \emph{semi-superluminal} if only one is.}

Suppose, in  a world $M_1$  a photon   is sent from the Earth
(we denote this event  $s_1$) to arrive at a distant star at some moment $\tau_1$  by the clock of that star.  Let, further, $M_2$ be the world which was initially the same as  $M_1$ (whether it \emph{was} the same may depend on what  theory  we are using for our analysis of the situation),
but in which instead of the photon a mighty spaceship is sent to the star (the start of the
spaceship is $s_2$).
 On its way to the star the spaceship warps and tears the spacetime by
exploding passing stars, merging binary black holes and triggering other imaginable powerful processes. Assuming that no superluminal (``tachyonic") matter is involved,
the spaceship arrives at the star later than the photon emitted in $s_2$, but nevertheless
it is imaginable that its arrival time  $\tau_2$ is \emph{less} than $\tau_1$. Thus, the speed of the spaceship in one world ($M_2$) would exceed the speed of light in another  ($M_1$), which would not contradict  the non-tachyonic nature of the spaceship.
Nor would such a flight break the ``light barrier" in $M_1$: the inequality $\tau_2<\tau_1$ does imply that the front  $\EuScript F_1$ is superluminal, but no  \emph{material signal} in $M_1$ corresponds to that front. In particular, there is  no spaceship in \emph{that} spacetime associated with $\EuScript F_1$.
It is such a pair of worlds
   $M_{1,2}$  that we call a semi-superluminal alternative. A theory admitting such alternatives allows superluminal signalling without tachyons.
\statmaph{Example}{\label{ex:hyperj}
Let $M_1$ be a Minkowski plane and $s_1$ be its point with the coordinates $t=-3/2$, $x=-1$. Let, further, $M_2$  be the spacetime obtained by removing the segments  $t\in[-1,1]$, $x=\pm 1$ from another Minkowski plane
\begin{figure}[t,b]
\begin{center}
\includegraphics[width=\textwidth]{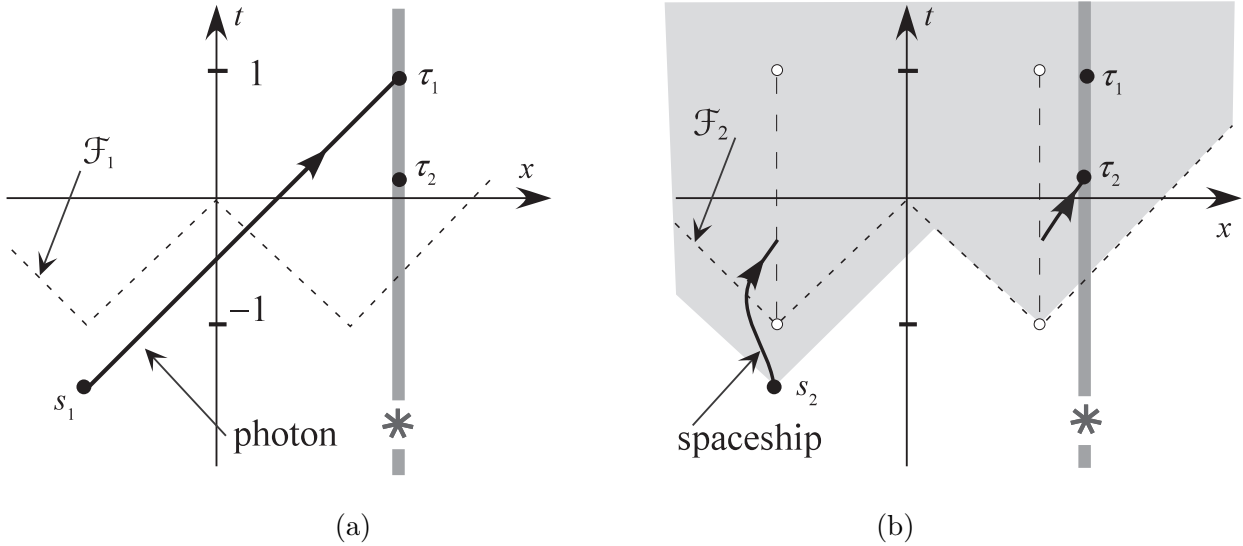}\\
 \hfill (a)\hspace{0.4\textwidth} (b)\hfill\hspace*{0pt}\\
\end{center}
 \caption{a) The world $M_1$. \label{fig:hyperj} b) The world $M_2$. The shaded region is the causal future of $s_2$, and the dashed broken line is the front  $\EuScript F_2$, which bounds $N_2^*$. }
\end{figure}
and gluing the left/right bank of either cut to the right/left bank of the other one.  The differences between $M_1$ and $M_2$ are confined, in a sense,  to the future of the points   $t=-1$, $x=-1$ and  $t=-1$, $x=1$,  see~figure~\ref{fig:hyperj}. Speaking more formally,
 $N^*_1$  is the complement to the union of two future cones with the vertices at those two points.
 That  $N^*_1$  is maximal indeed is clear from the fact that any larger past set   would contain
a past directed timelike curve $\lambda$ terminating at one of the mentioned vertices, while  $\isom(\lambda)$ cannot
have a past end point (because of the singularity).

Evidently,
$\EuScript F_1\not\subset \overline{J^+_{M_1}(s_1)}$, so $\EuScript F_1$ is superluminal. At the same time the surface $\EuScript F_1\subset M_1$
does not correspond to any signal  in $M_1$ (see above). And the front
$\EuScript F_2$   \emph{is not} superluminal, whence we conclude that the alternative $(M_k,\metr_k,s_k)$ is
  \emph{semi-}superluminal. Although the spaceship reaches the destination sooner than the photon shown in figure~\ref{fig:hyperj}, the photon belongs to another universe.  In its own universe $M_2$ the  spaceship moves on a timelike curve, in  full agreement with its non-tachyonic nature.}

A flaw in the just considered alternative is that the difference between $M_1$ and $M_2$ is too exotic.  One cannot say today whether ``the topology change" of that kind (if possible at all) can be attributable to something that takes place in $s_{1,2}$. Unfortunately, this is a general rule: as the following proposition shows, the spacetimes of a semi-superluminal alternative cannot be ``too nice".

\statmaph{Proposition}{\label{prop:hot1}
The spacetimes $M_1$ and $M_2$ of a semi-superluminal   alternative  $(M_k,\metr_k,s_k)$  cannot both be globally hyperbolic.}
\statmaph{Remark}{ Note the difference between this proposition and the statement proven earlier to the effect that within general relativity the spacetimes  of a \emph{superluminal}   alternative  cannot both be globally hyperbolic. The former, in contrast to the latter, states a purely kinematical fact depending neither on the Einstein equations, nor on  criteria of admissibility of alternatives. Essentially, that fact is just a geometrical property of globally hyperbolic spacetimes.}
\par
\begin{figure}
  \includegraphics[width=\textwidth]{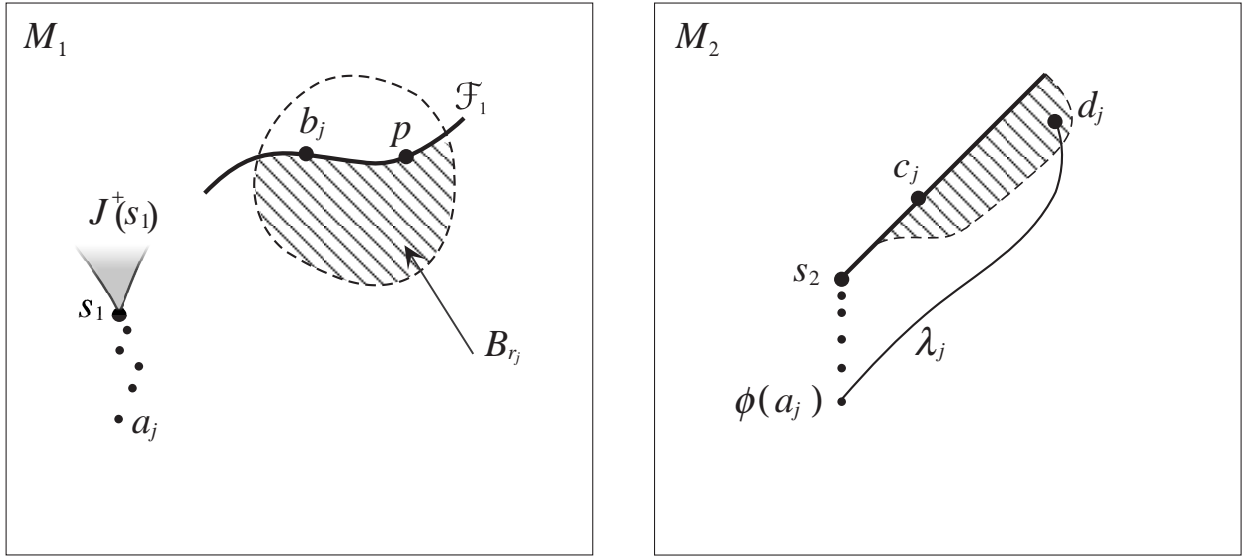}\\
  \caption{\label{fighy4}
The hatched regions are $N^*_1\cap B_{r_j}$ and its image under $\isom$, respectively. The ball $B_{r_j}$ bounded by the dashed line lies, by hypothesis, outside
  $J^+(s_1)$. But this contradicts the fact that the
 curves $\isom^{-1}(\lambda_j)$ must converge to a future directed
 curve from $s_1$ to $p$.}
\end{figure}

\noindent\emph{Proof.}
Suppose  that the front $\EuScript F_1$ is  superluminal.
Then   some of its points  must be separated from the --- closed by the global hyperbolicity of $M_1$, see proposition 6.6.1 of \cite{HawEl} --- set $J^+(s_1)$, that is there
must be a point $p$, see~figure~\ref{fighy4}, such that
\[
p\in \EuScript F_1, \qquad
B_r\cap\overline{J^+(s_1)}=\varnothing\quad \forall r<\bar r,
 \]
 where $\bar r$ is a constant, and $B_r$ is a coordinate ball of radius $r$  centered at $p$.

 Pick a sequence $ a_j\in I^-( s_1) $, $j=1,2,\ldots$ converging to  $ s_1$.
Our goal is to demonstrate that, unless $\EuScript F_2$ is superluminal,  there is a timelike curve $\mu_j$ from $a_j$ to $B_{r_j}$ for any $j$ and any $ r_j <\bar r$. That will prove the proposition, since   $r_j$ can be chosen so as to converge to zero. The future end points of $\mu_j$ in such a case will  converge to  $p$, which would  imply, by lemma 14.22 of \cite{neil}. that $p\in J^+(s_1)$ in contradiction to the choice of $p$.

To find for a given $j$ a curve $\mu_j$ of the just mentioned type, pick a pair of points
\[
b_j\in\bigl(\EuScript F_1\cap
B_{r_j}\bigr)\quad\text{and}\quad c_j\in \EuScript F_2,
\]
such that for any their neighbourhoods $U_j\supset b_j$ and $V _j\supset c_j$ it is true that
\[
\isom\bigl(N^*_1\cap
U_j\bigr) \cap V_j \neq \varnothing.
\label{eq:byb ck}\tag{$*$}
\]
To see that such pairs always exist, note that otherwise the maximal --- by hypothesis --- spacetime $M_2$, would have an extension $M_2^\text{ext}\equiv B_{r_j}\cup_{\isom'} M_2$, where $\isom'$ is the restriction of  $\isom$ to a connected component of  $N^*_1 \cap B_{r_j}$ (obviously, $M_2^\text{ext}$ is a smooth connected  pseudo-Riemannian  manifold containing $M_2$ as a proper subset. So, it is an  extension of $M_2$, if it is Hausdorff, i.~e. if there are no points $b_j$, $c_j$).

Now assume that  $\EuScript F_2$ is \emph{not} superluminal. Then $c_j$ being a point of $\EuScript F_2$ must lie in $\overline{J^+(s_2)}$, and hence  in the (closed) set $J^+(s_2)$ too.
 Thus (recall that $a_j\prec s_1$, whence   $\isom(a_j)\prec s_2$) a pair $a_j, c_j$ can be found such that
 \[
 \isom(a_j)\prec s_2\preccurlyeq c_j.
 \]
 By proposition 4.5.10 of \cite{HawEl} it follows that
$\isom(a_j)\prec c_j$. Hence there is a neighbourhood of $c_j$ which lies entirely in the open --- by \cite[lemma 14.3]{neil} --- set $I^+(\isom(a_j))$.
And according to
\eqref{eq:byb ck} that neighbourhood contains points of $\isom(N^*_1\cap
B_{r_j} )$. So there also must exist   points $d_j$:
\[
\isom(a_j)\prec d_j,\qquad d_j\in \isom(N^*_1\cap
B_{r_j} ) \subset N^*_2 .
\]

The last inclusion  coupled with the fact that
$N^*_2$ is a past set means that the timelike curve $\lambda_j$ connecting
$\isom(a_j)$ with $d_j$  lies entirely in $N^*_2$ and thus defines the curve
$\mu_j\equiv\isom^{-1}(\lambda_j)$. The latter possesses all the desired properties: it is timelike, it starts in $a_j$, and it ends in $B_{r_j}$.

\section*{Acknowledgment}
 This work was supported by RFBR Grant
No.~15-02-06818.

\end{document}